\documentclass[preprint,showpacs,preprintnumbers,superscriptaddress]{revtex4-1}

\usepackage{graphicx}
\usepackage{dcolumn}
\usepackage{amssymb}
\usepackage{mathrsfs}
\usepackage{amsmath}
\usepackage{epsfig}
\usepackage{hhline}
\usepackage{color}
\usepackage{multirow}
\usepackage[T1]{fontenc}
\usepackage{verbatim}
\usepackage{slashed}
\usepackage{hyperref}
\usepackage{diagbox}
\usepackage{makecell}

\begin{document}

\title{Search for nearly degenerate higgsinos via photon fusion with the semileptonic channel at the LHC}
\author{Hang Zhou}
\affiliation{School of Microelectronics and Control Engineering, Changzhou University, Changzhou, 213164, China\\
             Email: zhouhang@cczu.edu.cn}
\author{Ning Liu}
\affiliation{Physics Department and Institute of Theoretical Physics, Nanjing Normal University, Nanjing, 210023, China\\
             Email: liuning@njnu.edu.cn}

\begin{abstract}
Electroweak scale higgsinos with a nearly degenerate spectrum in supersymmetric models are well-motivated, but generally less constrained at collider experiments as the decay products are often too soft to detect. Initial photon fusions alongside the collision of protons at the Large Hadron Collider (LHC) have drawn attention recently as a way to search for new physics with such kind of spectra. In this paper, we demonstrate a search strategy for chargino pair production from photon fusion $pp\to p(\gamma\gamma\to\tilde{\chi}^{+}_{1}\tilde{\chi}^{-}_{1})p$ at the 13\,TeV LHC via the semileptonic decay channel, as a probe for the compressed spectra of higgsinos. Forward detectors make it possible to detect the outgoing protons after emitting the initial photons in these processes. We here provide simple event selections on missing energy and transverse momentum of leptons, which are effective enough to reach significant sensitivity. The chargino mass can be excluded at 95\% C.L. up to about 295\,GeV with the mass difference $\Delta m(\tilde{\chi}^{\pm}_{1},\tilde{\chi}^{0}_{1})$ being only a few GeV with the integrated luminosity of 3\,ab$^{-1}$. With a relatively small luminosity of 100\,fb$^{-1}$, the $2\sigma$ exclusion bounds can as well exceed current experimental limits in the range of $\Delta m=1\sim2$\,GeV, reaching over 190\,GeV for chargino mass.
\end{abstract}
\maketitle

\section{Introduction}

Supersymmetric models have long been perceived as one of the natural extensions of the Standard Model (SM) to explain the naturalness problem as well as other beyond SM phenomena, including the nature of the dark matter. Tremendous effort has been made for decades to search for the supersymmetric (SUSY) particles through a variety of experiments, especially at the Large Hadron Collider (LHC), yet with null results to date. Consequently, SUSY particles have been cornered in an ever narrower parametric space and the low-energy SUSY models are severely restricted. Based on full set of data during Run 2 at the LHC (with an integrated luminosity of 139 fb$^{-1}$ at a 13 TeV center-of-mass energy), recent analyses on the search for pair production of gluinos or squarks have excluded their masses up to 2.2 TeV and 1.7 TeV, respectively, assuming either R-parity conservation or violation~\cite{ATLAS:2023afl}. And the chargino mass can be excluded in the range from 260 to 520 GeV for a mass splitting between chargino and neutralino $\Delta m(\tilde{\chi}^{\pm}_{1},\tilde{\chi}^{0}_{1}$) around the electroweak scale, with a massless neutralino~\cite{ATLAS:2023act}. As the mass splitting gets smaller below 100 GeV, the mass bound for chargino ranges from 200 to 300 GeV within a simplified SUSY model~\cite{ATLAS:2021moa}. For an even more compressed spectra of the electroweakinos, the most constraint lower limit on chargino mass still comes from the LEP2 experiment around 90 GeV~\cite{LEPbound}. Rough patterns appear that the restrictions on the electroweakino masses get weaker with a smaller mass splitting.

This less restrictive corner of parameter space survives partly because of the low cross sections for charginos and neutralinos at the LHC as they cannot be produced through strong interaction. On the other hand, the compressed mass splitting between them often leads to soft leptons or jets in the final state that can hardly be detected and requires additional photons or jets from the initial state radiation to serve as a trigger for detection~\cite{Han:2013usa,Schwaller:2013baa,Baer:2014kya}. Last but not the least, difficulties in accurate reconstruction for neutralinos as a missing momentum also make the compressed scenario less constrained. Strategies have been proposed taking advantage of the disappearing tracks to probe such spectra with the mass splitting down to several hundreds of MeV~\cite{Fukuda:2017jmk,Mahbubani:2017gjh}. For the spectra not squeezed that much, it has been suggested that utilizing forward detectors such as CMS-TOTEM Precision Proton Spectrometer~\cite{CTPPS2014} or ATLAS Forward Proton~\cite{AFP2015}, the LHC can be used as a photon collider to probe light SUSY particles characteristic of nearly degeneracy~\cite{Beresford:2018pbt,Harland-Lang:2018hmi,Godunov:2019jib,Zhou:2022jgj}. Unlike the deep inelastic collisions studied in the traditional searches at the hadron collider, these proposals target the unsuccessful ones termed as the ultraperipheral collision, throughout which the colliding protons remain intact and can be detected by the above-mentioned forward detectors. What matters is that electromagnetic field around the high-velocity protons can be approximated to equivalent on-shell photons~\cite{Budnev:1975poe}, which may also collide with each other giving rise to photon fusions at the LHC
\begin{align}
\gamma\gamma\to X^{+Q}X^{-Q},
\label{upc}
\end{align}
where $X^{\pm Q}$ refer to particles carrying $Q$ units of electric charge. Detection of the outgoing unscathed protons can then help determine the 3-dimensional missing momentum in the final states, rather than the transverse missing one, realizing a more sensitive search. Further, the cross sections of such ultraperipheral collisions have been demonstrated to be large enough for promising search for new physics beyond the SM~\cite{Ohnemus:1993qw,Schul:2008sr,Fichet:2013gsa,Khoze:2017igg}.

As a complementary research to our previous work~\cite{Zhou:2022jgj} searching for the nearly degenerate higgsinos via photon fusion at the LHC utilizing the leptonic decay channel of the $W$ boson, we will demonstrate in this paper that the semileptonic channel can also reach significant sensitivity for such a compressed scenario of SUSY models. In the next section, we introduce the general parameter setting for the investigated scenario and relevant processes as our signal and background. Simulations and search strategies are presented in Section III. Section IV is the conclusion.

\section{The higgsino world scenario and signal at the LHC}

We consider the aforementioned compressed scenario within the framework of the Minimal Supersymmetric Standard Model (also known as the MSSM), which realizes SUSY in a way as simple as possible and at the same time, can be satisfactory with regard to the naturalness problem. In a quantitative way, minimizing the Higgs potential relates the mass of the SM $Z$ boson to the two Higgs doublets in the MSSM and the superpartner higgsino at tree level~\cite{Arnowitt:1992qp}
\begin{align}
\frac{m^{2}_{Z}}{2}=\frac{m^{2}_{H_{u}}\tan^{2}\beta-m^{2}_{H_{d}}}{1-\tan^{2}\beta}-\mu^{2},
\label{min_pot}
\end{align}
in which $\mu$ is the mass parameter of higgsino, $m_{H_{u}(H_{d})}$ is the soft SUSY breaking mass of the Higgs doublet $H_{u}$($H_{d}$). The ratio of their vacuum expected values is defined as $\tan\beta\equiv v_{u}/v_{d}$. A natural argument then arises for the terms of Eq.\eqref{min_pot} to be in a similar order of magnitudes, if we require an observed $Z$ mass and no significant fine-tuning in parameters~\cite{Barbieri:1987fn}. In consideration of the 125 GeV SM Higgs boson and current lower limits on top squark mass and gluino mass around 2 TeV, one of the commonly accepted schemes is called the higgsino world where superscalars are of order a few TeV while the higgsino masses are relatively low at sub-TeV~\cite{Kane:1997fa}. Such schemes include the minimal supergravity~\cite{Feng:1999mn} and the radiative nSUSY~\cite{Baer:2012up}, and the spectra of light higgsinos also lead to rich phenomenology related to dark matter search, muon $g-2$ and collider experiments~\cite{Hikasa:2015lma,Han:2020exx,Lv:2022pme,Abdughani:2017dqs}. Assuming a wide gap between $\mu$ and $M_{1,2,3}$, the masses for bino, wino and gluino respectively, the lightest mass eigenstates chargino and neutralino can then acquire a relatively tiny mass splitting of a few GeV with a relatively large $\tan\beta>10$~\cite{Giudice:1995np},
\begin{align}
\Delta m({\tilde{\chi}^{\pm}_{1},\tilde{\chi}^{0}_{1}})=\frac{m^{2}_{W}}{2M_{1}}\left(1-\sin2\beta-\frac{2\mu}{M_{2}}\right)+\frac{m^{2}_{W}}{2M_{2}}(1+\sin2\beta)\tan^{2}\theta_{W},
\end{align}
in which $\theta_{W}$ is the Weinberg angle and $m_{W}$ the SM $W$ mass. As discussed in Section I, this light and nearly degenerate spectrum of electroweakinos is less constrained by collider experiments and we use SUSYHIT package~\cite{Djouadi:2006bz} to generate such mass spectra which will then be used for event generation and further analysis.

\begin{figure}[t]
\centering
\begin{minipage}{0.38\linewidth}
  \centerline{\includegraphics[scale=0.67]{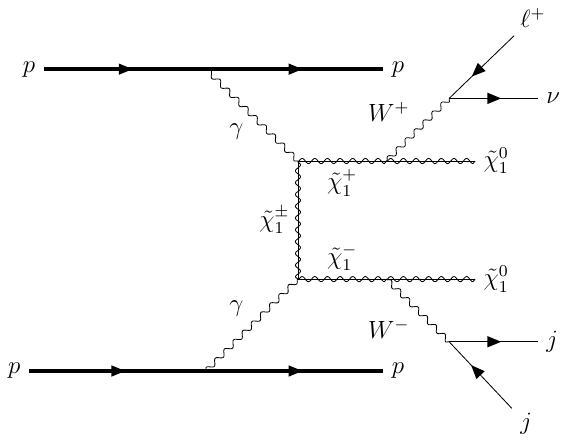}}
  \centerline{(a)}
\end{minipage}
\qquad\qquad
\begin{minipage}{0.38\linewidth}
  \centerline{\includegraphics[scale=0.67]{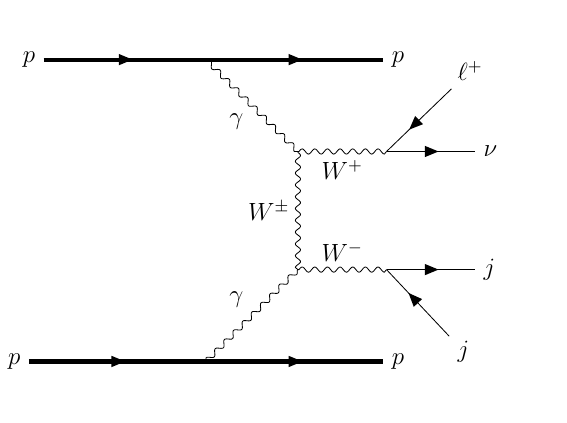}}
  \centerline{(b)}
\end{minipage}
\\[12pt]
\begin{minipage}{0.38\linewidth}
  \centerline{\includegraphics[scale=0.67]{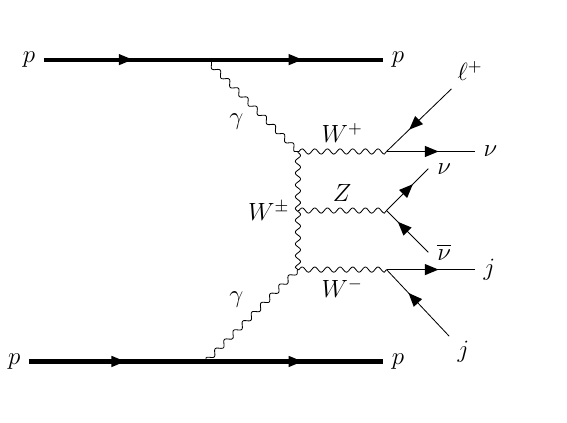}}
  \centerline{(c)}
\end{minipage}
\qquad\qquad
\begin{minipage}{0.38\linewidth}
  \centerline{\includegraphics[scale=0.67]{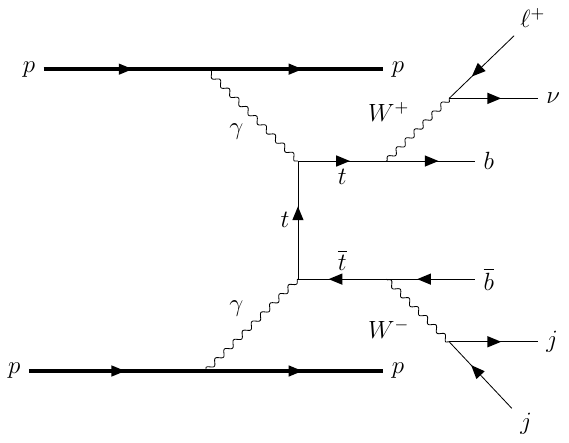}}
  \centerline{(d)}
\end{minipage}
\caption{Feynman diagrams of the signal and SM background of semileptonic channel search from chargino decay via equivalent photon fusion at the LHC. Corresponding Hermitian processes are not shown in the diagrams, but are taken into consideration in the following simulations and analysis.}
\label{fig:FD}
\end{figure}

In the pair production of charginos from photon fusion through the above-mentioned UPC (Eq.\eqref{upc}), virtual $W$ bosons can mediate the subsequent decays to neutralinos with either leptonic or hadronic final states, as the mass difference between the parent chargino and neutralino is tiny and real $W$ bosons cannot be produced in this compressed spectrum. We focus on the semileptonic case in the present paper as a complementary study to the fully leptonic one~\cite{Zhou:2022jgj}, the Feynman diagram of which as our signal is displayed in FIG.~\ref{fig:FD}(a):
\begin{align}
\label{signal}
pp\to p\left(\gamma\gamma\to\tilde{\chi}^{+}_{1}\tilde{\chi}^{-}_{1}\to\tilde{\chi}^{0}_{1}W^{+}\tilde{\chi}^{0}_{1}W^{-}\to\tilde{\chi}^{0}_{1}\tilde{\chi}^{0}_{1}+\ell^{\pm}\nu(\bar{\nu})+\text{jets}\right)p\,.
\end{align}
Final states consist of a large amount of missing momentum and soft leptons, as well as two intact outgoing protons which will be seen by the forward detectors. Similar signatures can be produced within the SM through the following processes,
\begin{align}
\label{bkg-1}
pp\to{}& p\left(\gamma\gamma\to W^{+}W^{-}\to\ell^{\pm}\nu(\bar{\nu})+\text{jets}\right)p\,,\\
\label{bkg-3}
pp\to{}& p\left(\gamma\gamma\to W^{+}W^{-}Z\to\ell^{\pm}\nu(\bar{\nu})+\text{jets}+\nu\bar{\nu}\right)p\,,\\
\label{bkg-4}
pp\to{}& p\left(\gamma\gamma\to t\bar{t}\to b\bar{b}+\ell^{\pm}\nu(\bar{\nu})+\text{jets}\right)p\,,
\end{align}
as shown in FIG.~\ref{fig:FD}\,(b)(c)(d). Feynman diagrams in the present paper are drawn using the TikZ-Feynman package~\cite{Ellis:2016jkw}. Note that for simplicity, the corresponding Hermitian conjugate processes are not shown in the Feynman diagrams, which will be included in the following simulations and further analysis. The most dominant background comes from the $WW$ events (Eq.~\eqref{bkg-1}), the cross section of which at the 13\,TeV LHC is of order 0.01\,pb. While for the other two subdominant ones (Eq.~\eqref{bkg-3} and \eqref{bkg-4}), their cross sections are both of order $10^{-5}$\,pb. Although the $WWZ$ background (Eq.~\eqref{bkg-3}) contributes as an irreducible one, the low production rates of this kind of three-boson events will not be a threat to the signal significance, which will be shown in the following section. And the $t\bar{t}$ events can be well suppressed via b-veto, apart from other kinematical cuts. It should be noted that there also exists the combinatorial background from high pile-up collisions, leading to fake protons in the forward detectors. Time-of-flight detectors can be used as effective ways tackling these issues, refer to \cite{Cerny:2020rvp} and other relevant studies on the combinatorial background \cite{Harland-Lang:2018hmi,Goncalves:2020saa,Martins:2022dfg,Harland-Lang:2022jwn}, which is beyond the scope of the present paper.

\begin{figure}[t]
\centering
\begin{minipage}{0.38\linewidth}
  \centerline{\includegraphics[scale=0.45]{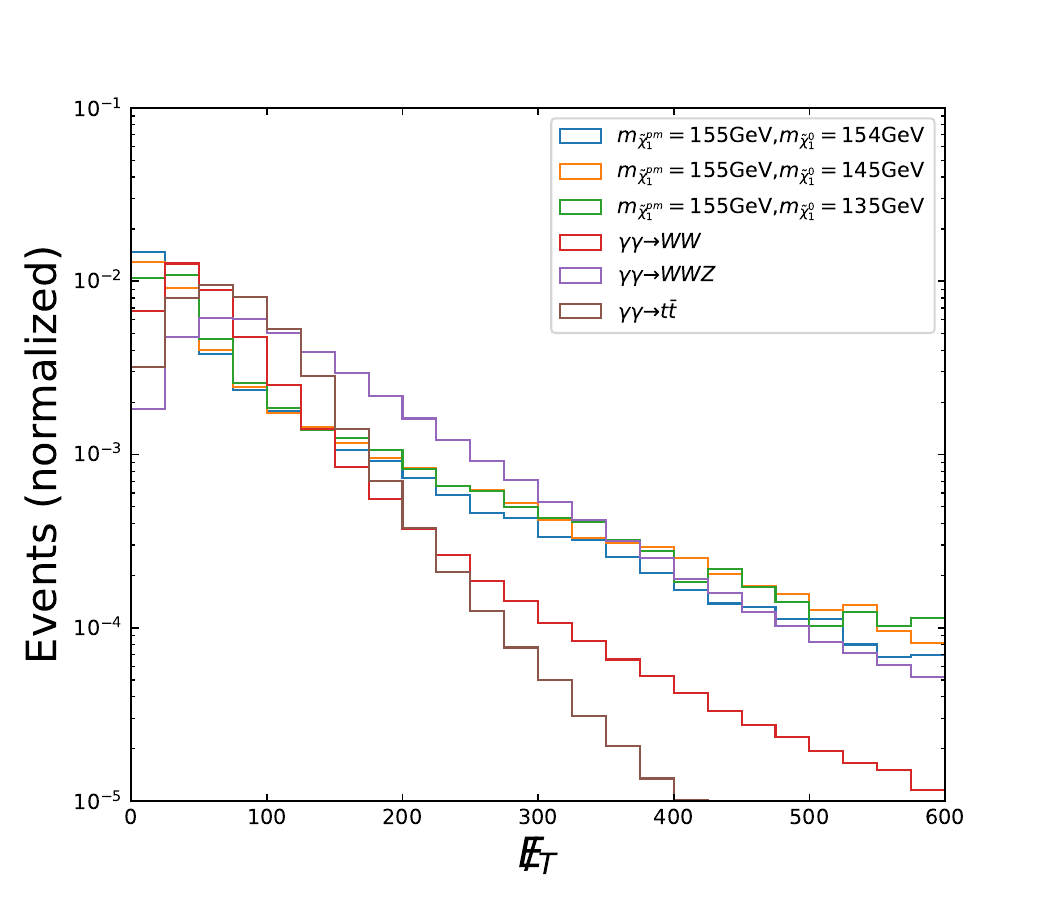}}
  \centerline{(a)}
\end{minipage}
\qquad\qquad
\begin{minipage}{0.38\linewidth}
  \centerline{\includegraphics[scale=0.45]{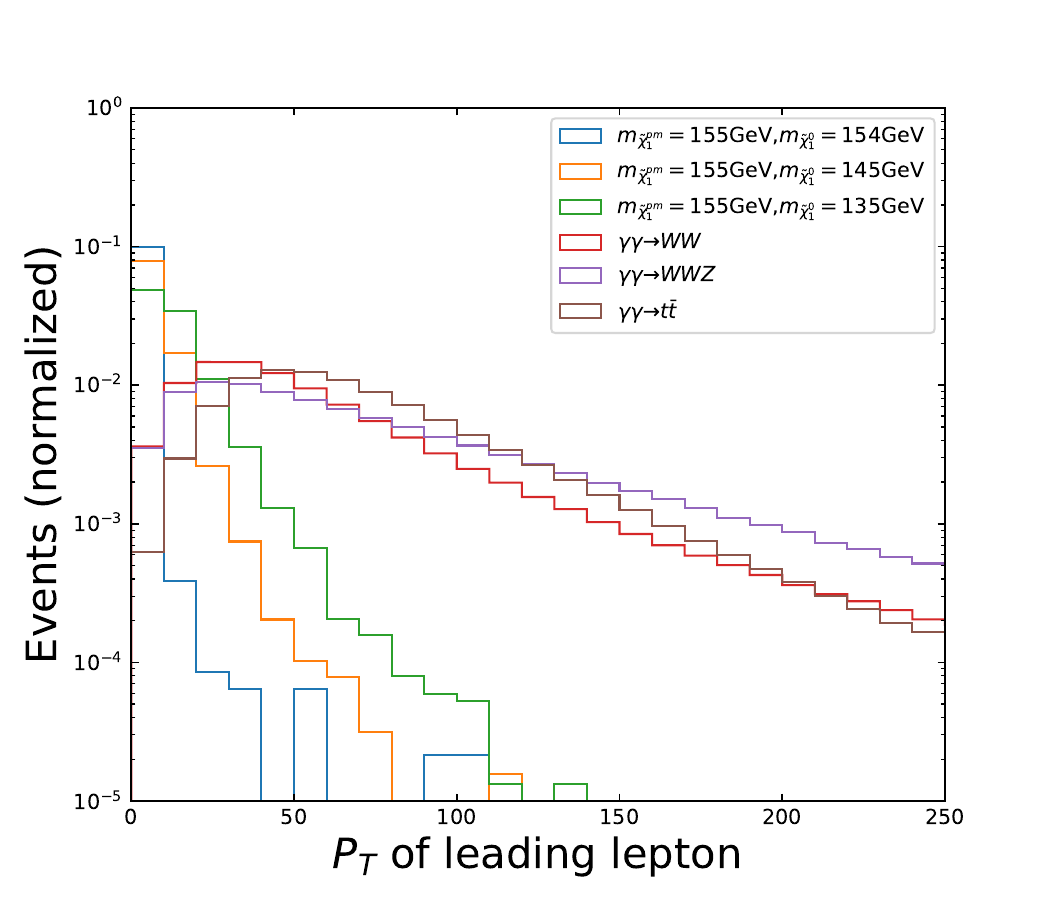}}
  \centerline{(b)}
\end{minipage}
\\[12pt]
\begin{minipage}{0.38\linewidth}
  \centerline{\includegraphics[scale=0.45]{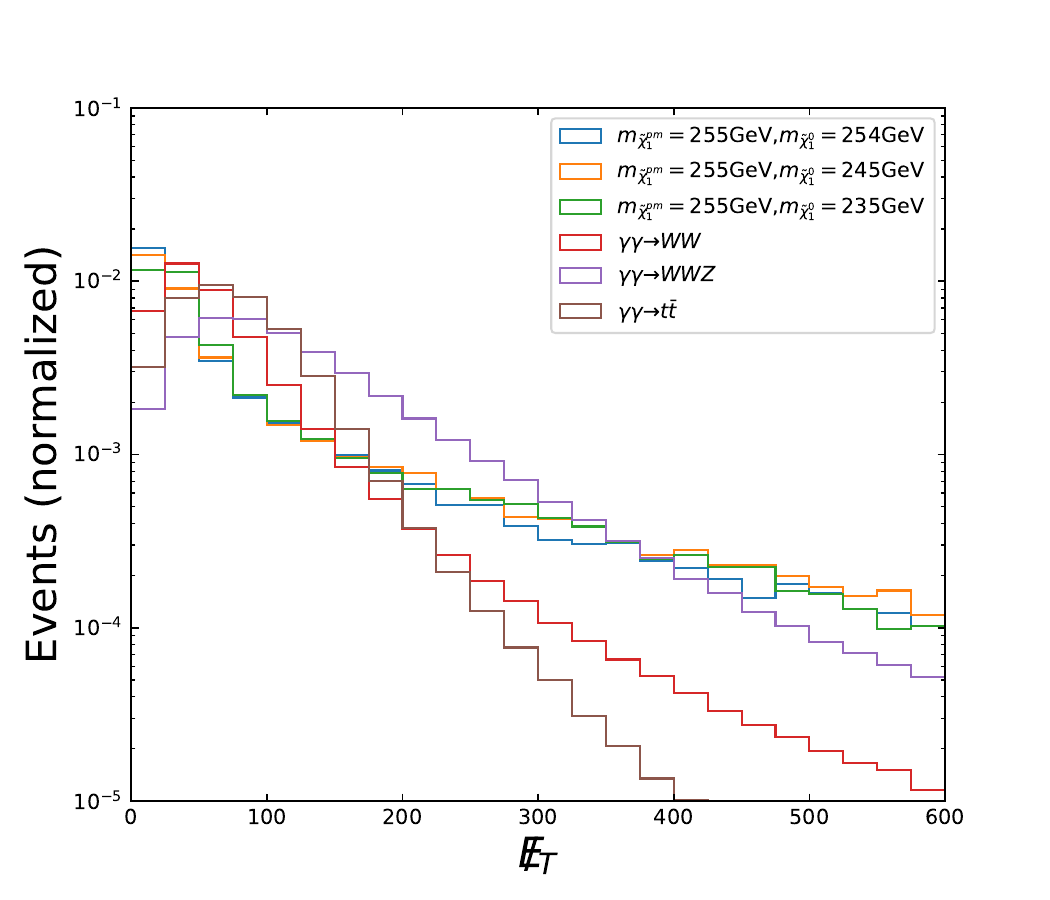}}
  \centerline{(c)}
\end{minipage}
\qquad\qquad
\begin{minipage}{0.38\linewidth}
  \centerline{\includegraphics[scale=0.45]{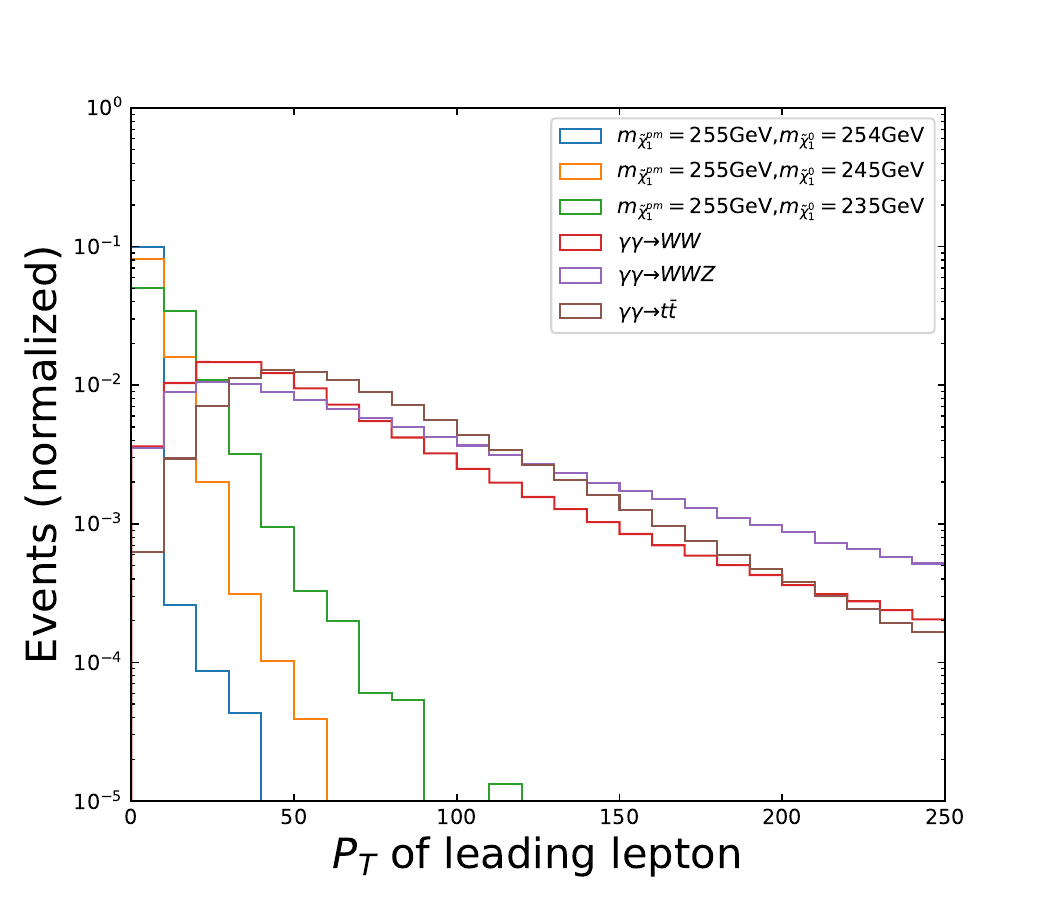}}
  \centerline{(d)}
\end{minipage}
\caption{Kinematic distributions of signal $pp\to p\left(\gamma\gamma\to\tilde{\chi}^{+}_{1}\tilde{\chi}^{-}_{1}\to\tilde{\chi}^{0}_{1}\tilde{\chi}^{0}_{1}+\ell^{\pm}\nu(\bar{\nu})+\text{jets}\right)p$ versus backgrounds including the $WW$, $WWZ$ and $t\bar{t}$ events from photon fusion at the 13 TeV LHC. The benchmarks are chosen as chargino mass $m_{\tilde{\chi}^{\pm}_{1}}=155$ and 255 GeV, with the mass difference between chargino and neutralino $\Delta m$=1, 10 and 20 GeV.}
\label{fig:dist}
\end{figure}

\section{Simulation and search strategies}

To arrive at the sensitivity of searching for the semileptonic decay channel from charginos Eq.~\eqref{signal} at the LHC, we perform simulations for both of the signal and the SM background. Parton-level events are generated via MG5. Parton showering and detector simulation are performed using \textsc{pythia 8.2} \cite{Sjostrand:2014zea} and \textsc{Delphes 3.5.0} \cite{deFavereau:2013fsa}. \textsc{checkmate2} (version 2.0) \cite{Dercks2017} is also used to connect the workflow of the above processes and for further analysis of the final-state kinematics and cutflow.

The signal and background differ in the most distinctive way in regard to the magnitude of missing momentum and the softness of final states, including leptons and jets. In the signal process, missing momentum comes from massive neutralinos while the leptons or jets, coming from off-shell $W$ bosons, keep only a small fraction of the total energy, being much more soft. While in the background process, the missing momentum is consisted of massless neutrinos (it is quite safe to assume the SM neutrinos to be massless in our simulation) and the final leptons or jets are decayed from much more energetic $W$ bosons. Distributions of missing transverse energy $\slashed{E}_{T}$ and transverse momentum of the leading leptons $p_{T}(\ell_{1})$ are presented in FIG.~\ref{fig:dist} to illustrate the kinematic difference between signal and background. The benchmarks are chosen as $m_{\tilde{\chi}^{\pm}_{1}}=$155 and 255 GeV, with $\Delta m=$1, 10 and 20 GeV. Simple patterns appear in the kinematic distributions, as expected for the above discussions on the features of signal and background. The histograms of missing transverse energy tend to center around smaller values for the SM background than that for the signal (FIG.~\ref{fig:dist}(a)(c)). And harder leptons directly decayed from $W$ bosons lead to peaks at larger values for the background than that for the signal (FIG.~\ref{fig:dist}(b)(d)).

\begin{table}
\begin{tabular}{|c|*{5}{c|}}
\hline
\,$E_{\gamma}$\,(GeV)\, & \,(0,100]\, & \,(100,120]\, & \,(120,150]\, & \,(150,400]\, & \,(400,$+\infty$)\, \\ \hline
Eff. & 0 & 50\% & 70\% & 90\% & 80\% \\ \hline
\hline
\end{tabular}
\caption{Acceptance rates for initial photons for difference ranges of energies, which are equivalent to tagging efficiencies for the outgoing protons corresponding to their energy losses \cite{CTPPS2014,AFP2015}.}
\label{tab:protonrate}
\end{table}

According to these kinematic distributions, event selection will then be performed to suppress the background and to achieve better significance. But before that, note that the outgoing protons are not to be detected at a 100\% efficiency by the forward detectors, but rather at certain acceptance rates ($\xi$), which mainly depend on the energy losses of the protons.
For the range of $\xi\in(0.015, 0.15)$, acceptance rates for outgoing protons can be close to 100\%~\cite{CTPPS2014,AFP2015}, corresponding to energies of equivalent photons with $E_{\gamma}\in(100, 1000)$\,GeV at the 13\,TeV LHC. Since phenomenological studies generally indicate relatively lower acceptances $\sim90\%$ for outgoing protons when they emit photons \cite{Khoze2002}, hence we in the present work adopt conservative values for the proton tagging efficiencies varied with their energy losses, as shown in TABLE.~\ref{tab:protonrate}. And we then apply these rates to every generated event, serving as a procedure of pre-selection to realize the effect of forward detectors tagging the outgoing protons with these energy-dependant efficiencies. This is done by checking the energies of the initial fusing photons emitted by the protons going through ultraperipheral collision, since the protons energy losses actually correspond to energies of the initial photons exactly. Four-momenta of protons are also smeared in advance using a 5\%-width Gaussian function to simulate the resolution of the forward detectors $\simeq5$ GeV \cite{AFP2015}. PYLHE packages \cite{pylhe} are used when processing and smearing photon momenta from parton-level events.

\begin{table}
\centering
\begin{tabular}{|c|*{6}{c|}}
\hline
\diagbox{Cuts}{SR} & $\Delta m=1$ & $\Delta m=2$ & $\Delta m=5$ & $\Delta m=10$ & $\Delta m=15$ & $\Delta m=20$ \\ \hline
Cut-1 & \multicolumn{6}{|c|}{b-veto} \\ \hline
Cut-2 & \multicolumn{6}{|c|}{$\slashed{E}_{T}>10$} \\ \hline
Cut-3 & \,$P_{T}(\ell_{1})<2$\, & \,$P_{T}(\ell_{1})<3$\, & \,$P_{T}(\ell_{1})<5$\, & \,$P_{T}(\ell_{1})<7$\, & \,$P_{T}(\ell_{1})<9$\, & \,$P_{T}(\ell_{1})<12$\, \\ \hline
\hline
\end{tabular}
\caption{Cuts applied to proton-tagged events in six different signal regions divided according to mass splitting between chargino and neutralino. The masses, energies and momenta in the table are in units of GeV.}
\label{tab:SRcuts}
\end{table}

After the pre-selection, the surviving events are ones that are detected by the forward detectors, which will then go through two further cuts on missing transverse energy and on transverse momentum of the leading lepton. These cuts are applied based on their distributions for benchmarks with nearly degenerate chargino and neutralino of hundreds GeV scale, that is, chargino mass $m_{\tilde{\chi}^{\pm}_{1}}\in[105,275]$ GeV and mass splitting with neutralino $\Delta m=1,2,5,10,15,20$ GeV (Some of these benchmark points are shown for illustration in FIG.~\ref{fig:dist}). To realize better significance in this parametric space of interest, we perform the cutflow in six signal regions according to different mass splittings, as presented in TABLE.~\ref{tab:SRcuts}. Cut-1 rejects events with b-tagged jets identified to mainly suppress $t\bar{t}$ events. Cut-2 keeps events only with missing transverse energy larger than 10 GeV. Cut-3 requires small transverse momentum $P_{T}$ of leading lepton according to different $\Delta m$. At each step of event selection, the effective cross sections are also calculated for both signal and background, shown in TABLE.~\ref{tab:cutflow} as an example for one of the above benchmarks ($\Delta m$=5 GeV and $m_{\tilde{\chi}^{\pm}_{1}}=205$ GeV). The cutflow includes as well the pre-selection as tagging two outgoing protons according to tagging rates in TABLE.~\ref{tab:protonrate}. It can be seen from the cutflow that the most effective cut on the background events is requiring a relatively small $P_{T}$ for the leading leptons, since the energies of the final leptons in the signal come from a tiny mass splitting between chargino and neutralino. But for the background, there is no such limitation and the curves for distributions of $P_{T}(\ell_{1})$ slope more gently than that of the signal.

\begin{table}
\centering
\begin{tabular}{|l|c|c|c|c|}
\hline
  & \quad\makecell{$\Delta m$ = 5 GeV,\quad\quad \\ $m_{\tilde{\chi}^{\pm}_{1}}=205$ GeV\quad\quad} & \quad $W^{+}W^{-}$ \quad\quad & $WWZ$ & $t\bar{t}$ \\ \hline
No cuts applied & $6.480\times10^{-5}$ & $1.122\times10^{-2}$ & $5.861\times10^{-6}$ & $5.975\times10^{-6}$ \\ \hline
Outgoing protons tagged & $3.217\times10^{-5}$  & $2.162\times10^{-3}$ & $3.151\times10^{-6}$ & $2.716\times10^{-6}$ \\ \hline
Cut-1: b-veto & $3.140\times10^{-5}$ & $1.987\times10^{-3}$ & $2.901\times10^{-6}$ & $4.255\times10^{-7}$ \\ \hline
Cut-2: $\slashed{E}_{T}>10$\,GeV & $2.767\times10^{-5}$ & $1.933\times10^{-3}$ & $2.881\times10^{-6}$ & $4.202\times10^{-7}$ \\ \hline
Cut-3: $P_{T}(\ell_{1})<5$\,GeV & $9.924\times10^{-6}$ & $2.154\times10^{-5}$ & $2.890\times10^{-8}$ & $7.230\times10^{-10}$\\
\hline
\end{tabular}
\caption{Effective cross sections after each step of cutflow listed in TABLE.~\ref{tab:SRcuts} for signal in the signal region of $\Delta m$=5 GeV and $m_{\tilde{\chi}^{\pm}_{1}}=205$ GeV, as well as for the three SM backgrounds $W^{+}W^{-}$, $WWZ$ and $t\bar{t}$ events from photon fusion. Cross sections in the table are in units of picobarn.}
\label{tab:cutflow}
\end{table}

\begin{figure}[t]
\centering
\includegraphics{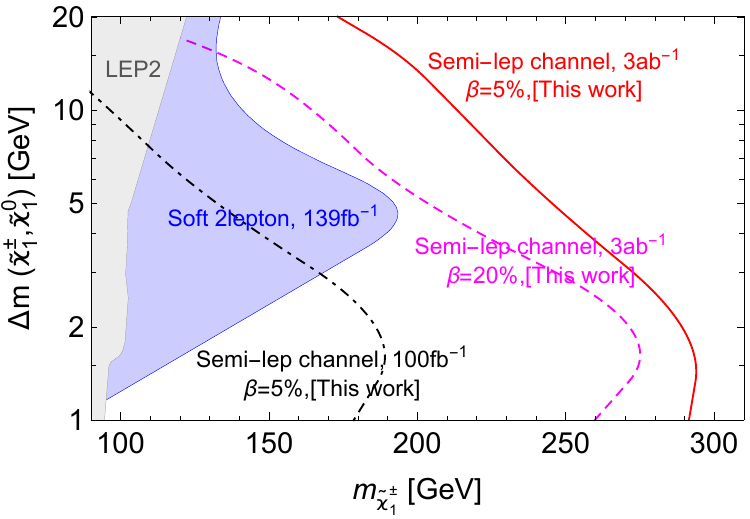}
\caption{$2\sigma$ exclusion limits on chargino mass and the mass difference $\Delta m$ with neutralino, based on a simulated search for the semileptonic channel via initial photon fusion at the 13\,TeV LHC.}
\label{fig:contour}
\end{figure}

Finally, statistical significance is calculated according to the formula $\alpha=S/\sqrt{B+(\beta B)^{2}}$, where $S$ ($B$) is the events number of signal (background) after the cutflow in TABLE.~\ref{tab:SRcuts}. Integrated luminosity is adopted as 100\,fb$^{-1}$ and 3\,ab$^{-1}$, and the systematic uncertainty is taken as $\beta=5\%$ and 20\%. $2\sigma$ exclusion limits are then obtained corresponding to luminosities of $\mathcal{L}=100\,\text{fb}^{-1}$ (black solid line) and $\mathcal{L}=3\,\text{ab}^{-1}$ (red and magenta solid lines) in FIG.~\ref{fig:contour}. As comparison, we present the current bounds from collider experiments for this space of parameter, including the blue region of the ATLAS search for soft final leptons at $\mathcal{L}=139\,\text{fb}^{-1}$ \cite{Aaboud2018} and the gray part from LEP2 experiment \cite{Heister2002}. From the contours of these exclusion bounds, the present work demonstrates that the search for the semileptonic channel can exceed limits given by current experiments, reaching about $m_{\tilde{\chi}^{\pm}_{1}}~190$\,GeV with the mass splitting of only $\sim1$\,GeV for $\mathcal{L}=100\,\text{fb}^{-1}$ (with $\beta=5\%$, black dot-dashed line). If the integrated luminosity is increased to 3\,ab$^{-1}$ ($\beta=5\%$), chargino mass can be excluded over 290\,GeV, at most 294\,GeV, for $\Delta m$ less than 2 GeV (red solid line). In the range of a slightly large mass difference of up to 20\,GeV, the $2\sigma$ exclusion limits on chargino mass can also reach $\sim175$\,GeV. Even if we take a conservative value of the systematic uncertainty as 20\%, the exclusion limit for $m_{\tilde{\chi}^{\pm}_{1}}$ can still reach 275\,GeV for $\Delta m$ of $1\sim2$\,GeV (magenta dashed line).

\section{Conclusion}
We demonstrate in this paper a search strategy for photon fusion to chargino pair at the 13\,TeV LHC via the semileptonic decay channel, as a probe for the nearly degenerate spectra of higgsinos. Colliding protons remain intact in these processes and can be detected by forward detectors like the AFP and CT-PPS at CERN. Simple event selections on missing energy and transverse momentum of leptons are effective enough to reach significant sensitivity. The chargino mass can be excluded at 95\%C.L. up to around 295\,GeV with the mass difference $\Delta m(\tilde{\chi}^{\pm}_{1},\tilde{\chi}^{0}_{1})$ being only a few GeV under the integrated luminosity of 3\,ab$^{-1}$ ($\beta=5\%$). With a relatively small luminosity of 100\,fb$^{-1}$, the $2\sigma$ exclusion bounds can also exceed current experimental limits in the range of $\Delta m=1\sim2$\,GeV, reaching over 190\,GeV for chargino mass.

\section{Acknowledgments}
This work is supported by the Natural Science Foundation of Jiangsu Province under Grant No. BK20230623, and the Natural Science Foundation of the Jiangsu Higher Education Institutions of China under Grant No.\,22KJB140007.


\begin{thebibliography}{99}

\bibitem{ATLAS:2023afl}
G.~Aad \textit{et al.} [ATLAS],
JHEP \textbf{02}, 107 (2024)

\bibitem{ATLAS:2023act}
G.~Aad \textit{et al.} [ATLAS],
JHEP \textbf{12}, 167 (2023)

\bibitem{ATLAS:2021moa}
G.~Aad \textit{et al.} [ATLAS],
Eur. Phys. J. C \textbf{81}, no.12, 1118 (2021)

\bibitem{LEPbound}
http://lepsusy.web.cern.ch/lepsusy/

\bibitem{Han:2013usa}
C.~Han, A.~Kobakhidze, N.~Liu, A.~Saavedra, L.~Wu and J.~M.~Yang,
JHEP \textbf{02}, 049 (2014)

\bibitem{Schwaller:2013baa}
P.~Schwaller and J.~Zurita,
JHEP \textbf{03}, 060 (2014)

\bibitem{Baer:2014kya}
H.~Baer, A.~Mustafayev and X.~Tata,
Phys. Rev. D \textbf{90}, no.11, 115007 (2014)

\bibitem{Fukuda:2017jmk}
H.~Fukuda, N.~Nagata, H.~Otono and S.~Shirai,
Phys. Lett. B \textbf{781}, 306-311 (2018)

\bibitem{Mahbubani:2017gjh}
R.~Mahbubani, P.~Schwaller and J.~Zurita,
JHEP \textbf{06}, 119 (2017)
[erratum: JHEP \textbf{10}, 061 (2017)]

\bibitem{CTPPS2014}
M.~Albrow \textit{et al.} [CMS and TOTEM],
CERN-LHCC-2014-021

\bibitem{AFP2015}
L.~Adamczyk, E.~Bana\'s, A.~Brandt, M.~Bruschi, S.~Grinstein, J.~Lange, M.~Rijssenbeek, P.~Sicho, R.~Staszewski and T.~Sykora, \textit{et al.}
CERN-LHCC-2015-009

\bibitem{Beresford:2018pbt}
L.~Beresford and J.~Liu,
Phys. Rev. Lett. \textbf{123}, no.14, 141801 (2019)

\bibitem{Harland-Lang:2018hmi}
L.~A.~Harland-Lang, V.~A.~Khoze, M.~G.~Ryskin and M.~Tasevsky,
JHEP \textbf{04}, 010 (2019)

\bibitem{Godunov:2019jib}
S.~I.~Godunov, V.~A.~Novikov, A.~N.~Rozanov, M.~I.~Vysotsky and E.~V.~Zhemchugov,
JHEP \textbf{01}, 143 (2020)

\bibitem{Zhou:2022jgj}
H.~Zhou and N.~Liu,
JHEP \textbf{10}, 092 (2022)

\bibitem{Budnev:1975poe}
V.~M.~Budnev, I.~F.~Ginzburg, G.~V.~Meledin and V.~G.~Serbo,
Phys. Rept. \textbf{15}, 181-281 (1975)

\bibitem{Ohnemus:1993qw}
J.~Ohnemus, T.~F.~Walsh and P.~M.~Zerwas,
Phys. Lett. B \textbf{328}, 369-373 (1994)

\bibitem{Schul:2008sr}
N.~Schul and K.~Piotrzkowski,
Nucl. Phys. B Proc. Suppl. \textbf{179-180}, 289-297 (2008)

\bibitem{Fichet:2013gsa}
S.~Fichet, G.~von Gersdorff, O.~Kepka, B.~Lenzi, C.~Royon and M.~Saimpert,
Phys. Rev. D \textbf{89}, 114004 (2014)

\bibitem{Khoze:2017igg}
V.~A.~Khoze, A.~D.~Martin and M.~G.~Ryskin,
J. Phys. G \textbf{44}, no.5, 055002 (2017)

\bibitem{Arnowitt:1992qp}
R.~L.~Arnowitt and P.~Nath,
Phys. Rev. D \textbf{46}, 3981-3986 (1992)

\bibitem{Barbieri:1987fn}
R.~Barbieri and G.~F.~Giudice,
Nucl. Phys. B \textbf{306}, 63-76 (1988)

\bibitem{Kane:1997fa}
G.~L.~Kane,
Nucl. Phys. B Proc. Suppl. \textbf{62}, 144-151 (1998)

\bibitem{Feng:1999mn}
J.~L.~Feng, K.~T.~Matchev and T.~Moroi,
Phys. Rev. Lett. \textbf{84}, 2322-2325 (2000)

\bibitem{Baer:2012up}
H.~Baer, V.~Barger, P.~Huang, A.~Mustafayev and X.~Tata,
Phys. Rev. Lett. \textbf{109}, 161802 (2012)

\bibitem{Hikasa:2015lma}
K.~i.~Hikasa, J.~Li, L.~Wu and J.~M.~Yang,
Phys. Rev. D \textbf{93}, no.3, 035003 (2016)

\bibitem{Han:2020exx}
C.~Han, M.~L.~L\'opez-Ib\'a\~nez, A.~Melis, \'O.~Vives, L.~Wu and J.~M.~Yang,
JHEP \textbf{05}, 102 (2020)

\bibitem{Lv:2022pme}
H.~Lv, D.~Wang and L.~Wu,
Phys. Rev. D \textbf{106}, no.5, 055008 (2022)

\bibitem{Abdughani:2017dqs}
M.~Abdughani, L.~Wu and J.~M.~Yang,
Eur. Phys. J. C \textbf{78}, no.1, 4 (2018)

\bibitem{Giudice:1995np}
G.~F.~Giudice and A.~Pomarol,
Phys. Lett. B \textbf{372}, 253-258 (1996)

\bibitem{Djouadi:2006bz}
A.~Djouadi, M.~M.~Muhlleitner and M.~Spira,
Acta Phys. Polon. B \textbf{38}, 635-644 (2007)

\bibitem{Ellis:2016jkw}
J.~Ellis,
Comput. Phys. Commun. \textbf{210}, 103-123 (2017)


\bibitem{Cerny:2020rvp}
K.~\v{C}ern\'y, T.~S\'ykora, M.~Ta\v{s}evsk\'y and R.~\v{Z}leb\v{c}\'\i{}k,
JINST \textbf{16}, no.01, P01030 (2021)

\bibitem{Goncalves:2020saa}
V.~P.~Gon\c{c}alves, D.~E.~Martins, M.~S.~Rangel and M.~Tasevsky,
Phys. Rev. D \textbf{102}, no.7, 074014 (2020)

\bibitem{Martins:2022dfg}
D.~E.~Martins, M.~Tasevsky and V.~P.~Goncalves,
Phys. Rev. D \textbf{105}, no.11, 114002 (2022)

\bibitem{Harland-Lang:2022jwn}
L.~A.~Harland-Lang and M.~Tasevsky,
Phys. Rev. D \textbf{107}, no.3, 033001 (2023)


\bibitem{Sjostrand:2014zea}
T.~Sj\"ostrand, S.~Ask, J.~R.~Christiansen, R.~Corke, N.~Desai, P.~Ilten, S.~Mrenna, S.~Prestel, C.~O.~Rasmussen and P.~Z.~Skands,
Comput. Phys. Commun. \textbf{191}, 159-177 (2015)


\bibitem{deFavereau:2013fsa}
J.~de Favereau \textit{et al.} [DELPHES 3],
JHEP \textbf{02}, 057 (2014)


\bibitem{Dercks2017}
D.~Dercks, N.~Desai, J.~S.~Kim, K.~Rolbiecki, J.~Tattersall and T.~Weber,
Comput. Phys. Commun. \textbf{221}, 383-418 (2017)

\bibitem{Khoze2002}
V.~A.~Khoze, A.~D.~Martin and M.~G.~Ryskin,
Eur. Phys. J. C \textbf{23}, 311-327 (2002)


\bibitem{pylhe}
Lukas Heinrich, Matthew Feickert, and Eduardo Rodrigues.
\newblock {pylhe: v0.3.0}\\
https://github.com/scikit-hep/pylhe/tree/v0.3.0

\bibitem{Aaboud2018}
M.~Aaboud \textit{et al.} [ATLAS],
Phys. Rev. D \textbf{97}, no.5, 052010 (2018)

\bibitem{Heister2002}
A.~Heister \textit{et al.} [ALEPH],
Phys. Lett. B \textbf{533}, 223-236 (2002)


\end{thebibliography}

\end{document}